\documentclass[usegraphicx,usenatbib]{mn2e}
\usepackage{times}
\def\mpc{h^{-1}{\rm{Mpc}}}
\def\apj {ApJ}

\def\aj {AJ}
\def\aap {A\&A}
\def\mnras {MNRAS}
\def\pasp {PASP}
\def\nn {astro-ph/}
\begin{document}
\title
[Alignments of Luminous Red Galaxies]
{Alignment between Luminous Red Galaxies and surrounding structures at $z\sim ~0.5$}
\author[Donoso et al.]
{Emilio Donoso$^{1,2}$\thanks{E-mail: edonoso@speedy.com.ar}, Ana O'Mill $^{3}$ \&
Diego G. Lambas$^{3,4}$  \\
$^1$Universidad Nacional de San Juan, Av. Ignacio de la Roza 590(O), 5400 San Juan, Argentina\\
$^2$Observatorio Astronomico Felix Aguilar, Av. Benavidez 8175(O), 5413 San Juan, Argentina\\
$^3$IATE, Observatorio Astronomico de la Universidad Nacional de Cordoba, Laprida 851, 5000 Cordoba, Argentina\\
$^4$Consejo de Investigaciones Cient\'{\i}ficas y T\'ecnicas (CONICET),
Avenida Rivadavia 1917, C1033AAJ, Buenos Aires, Argentina\\
}
\date{\today}
\pagerange{\pageref{firstpage}--\pageref{lastpage}} 
\maketitle
\label{firstpage}
\begin{abstract}
We analyse a high redshift sample ($0.4<z<0.5$) of LRG's extracted from the 
Sloan Digital Sky Survey Data Release 4 and their
 surrounding structures to explore the presence of alignment 
effects of these bright galaxies with neighbor objects.
 In order to avoid projection effects we compute photometric redshifts 
for galaxies within $3~\mpc$ in projection of LRGs and calculate 
the relative angle between the LRG major axis and the direction to 
neighbors within $1000$ km/s. 
We find a clear signal of alignment between LRG orientations 
and the distribution of galaxies within $1.5\mpc$. 
The alignment effects are present only for the red population
of tracers, LRG orientation is uncorrelated
 to the blue population of neighbor galaxies. 
These results add evidence to the alignment effects
 between primaries and satellites detected at low redshifts. We conclude that
such alignments were already present at $z\sim 0.5$.

\end{abstract}


\section{Introduction}

The study of galaxy orientations has the potential to provide important information
about the formation and evolution of cosmic structures. However, the interpretation of alignments,
and particularly the search of its observational evidence, has a long confusing history. 
Bingelli (1982) introduced the idea of analyzing the distribution of orientations of clusters
or their dominant galaxies relative to neighboring clusters or galaxies. Using a sample of well known
Abel clusters he found that the position angle of first ranked galaxies strongly correlates
with the orientation of the cluster itself, as determined from the spatial distribution of members. 
This author found that the orientation of a cluster is related to the distribution of 
neighboring clusters, i.e. clusters separated less than $\sim 30 \mpc$ tend to point to each other. 
Later, Struble and Peebles (1985) argued against this last effect claiming that it could be biased by
systematic errors. In agreement with Bingelli (1982), Argyres et al. (1986) found that galaxy counts are 
systematically high along the line defined by the major axis of Abell clusters or its dominant member,
up to at least $15 \mpc$ in projected distance.

Lambas et al. (1988) found that elliptical galaxies in the Uppsala General Catalog of Bright Galaxies
showed a significant alignment signal up to $2~\mpc$ with respect to the surrounding galaxies identified
by means of the Lick maps of galaxy counts, an effect likely to be associated to anisotropy of large
scales transferred to the elliptical galaxy population through mergers events. Since no similar
correlation was found for spiral galaxies, this gave rise to the first detection of a
morphology-orientation effect. Again, Rhee \& Katgert (1987) and West (1989) found convincing
evidence of Bingelli's results, and with a large sample of rich Abell clusters, Lambas et al. (1990) 
obtained a $30\%$ excess of brightest cluster galaxies pointing to the nearest-neighbor 
cluster at scales up to $15\mpc$. Also west (1989) presented evidence suggesting 
that the orientation of groups of galaxies in superclusters is not random, showing 
a strong correlation with surrounding groups within $\sim (15-30)\mpc$.

More recently, such alignment effects have also been found by Yang et al. (2006), 
using a large sample of $\sim 53.000$ poor galaxy groups extracted from the New York University
Value Added Galaxy Catalogue in the redshift range $0.01<z<0.2$. These authors determined that the 
distribution of satellite is aligned with the major axis of the brightest galaxies in groups. 
It is interesting to note that this systematic effect is statistically  significant only for the red 
population of satellites and  central galaxies. 

From the theoretical point of view, numerical studies (West et al. 1991; van Haarlem \& van de 
Weygaert 1993, Splinter et al.
1998; Faltenbacher et al. 2002) have shown that these cluster-cluster and substructure-cluster alignments
occur naturally in hierarchical clustering scenarios for structure formation such as the Cold Dark Matter Model;
a fact that can be interpreted as the result of correlations of density fluctuations at different
scales. 
Also, if galaxies formed after the collapse of their parent cluster, then the
anisotropic initial conditions could be imprinted in member galaxy orientations, producing alignment effects.
Both the tidal field of the parent cluster
(Barnes \& Efstathiou 1987, Usami \& Fujimoto 1997), or an anisotropic merger scenario,
(where interactions occur along the spatial directions corresponding to the primordial large-scale filaments)
could explain the strong alignment between dominant galaxies and surrounding structures. 

As previously discussed, evidence for the tendency of structures to align each other extends 
smoothly from the richest clusters of galaxies, to small groups and  
galaxies, over scales from $Kpc$ up to several $Mpc$, 
providing a test for models of the 
origin and evolution of structure in the universe. All these works concern 
low redshifts, typically $z\la 0.1$, and concentrate in rich and/or poor 
cluster environments. This limitation is not surprising since only a handful 
of high redshift clusters are known. 
Therefore, it is of interest to explore their presence in deeper samples. 

The Luminous Red Galaxies (LRG) sample corresponds mostly to intermediate redshift 
($0.2 \la z \la 0.55$) early type galaxies and, consequently, these objects may be ideal 
targets to test the alignment hypothesis at larger redshifts. A problem one faces 
with such a project is the lack of redshifts for the neighbor objects in projection 
that can dilute the alignment signals due mainly to contamination by distant 
background galaxies. In this paper, we use photometric redshifts calibrated 
with the publicly Artificial Neural Network code (Collister \& Lahav (2004), ANNz). The galaxy 
training set used in the code from galaxy sample randomly selected the SDSS DR4 
(main galaxy sample and LRGs). In this particular data set the $rms$ redshift 
error in the range $z_s<0.3$ is $\sigma_{rms}=0.03$ (O'Mill et al. 2006, submitted) 
to estimate the redshifts of objects close in projection to the 
LRGs and compute the relative angle between the LRG position angle and the 
radius vector from the LRG position to that of each neighbor galaxy. 
The data and statistics are briefly described as well as the main results and 
conclusions.

\section{Data}
The spectroscopic sample of the LRG used in this paper comprise the 
highest redshift objects of the total LRG sample. These galaxies were selected on the 
basis of color and magnitude to yield a sample of luminous intrinsically red 
galaxies (Eisenstein et al (2001)) that extends fainter than the SDSS galaxy spectroscopic 
sample. 

Together, this sample of luminous red galaxies covers an enormous volume of space, 
about $1h^{-3}$ Gpc$^3$ when completed, and is expected to trace clusters of galaxies 
at higher distances while providing a fairly homogeneous population of galaxies 
suitable to study large-scale structure formation and giant elliptical evolution. 
The details of selection algorithms as well as its tune-up to remove the passive 
evolution of an old stellar population are explained in Einsenstein (2001).

SDSS photometry is accurate to $\sim 2\%$ $rms$ in $g$,$r$,$i$ bands; 
and $\sim3\%$ $rms$ in $u$ and $z$. In the $r$ band, a $95\%$ 
completeness is achieved at $m_r\leq 22.2$. All magnitudes used in 
this work are modified Petrosian magnitudes (Blanton et al. (2001)), which measure galaxy 
fluxes within a circular aperture whose radius is determined by the azimuthally averaged 
light profile, and therefore measure a constant fraction of the total light independent 
of the position and distance. Spectroscopy is taken with two multifiber spectrographs 
that allow 640 spectra to be simultaneously acquired using pre-drilled aluminum plates. 
At a spectral resolution of 1800, the typical signal-to-noise per pixel is $>4$ at the 
peek of the $g$ band. Redshift accuracy is $30$ $km/s$. The fourth data release (DR4) 
covers a footprint area of 6670 and 4783 square degrees of sky for imaging and 
spectroscopy, respectively. A complete description of the survey is given 
by York et al. (2000).

Since we are interested in searching for alignment effects at the highest redshifts possible, 
we adopt the redshift range $0.4 <z <0.5$ for the LRG targets. Position angle $\theta$,
 major axis $a$ and minor axis $b$ of LRGs were derived by the $photo$ reduction 
pipeline from the $r$ band isophote at 25 magnitudes per square arcsec. Briefly 
explained, the radius of a particular isophote is measured as a function of angle 
and then expanded in Fourier series. Coefficients for this expansion are 
straightforward to calculate and account for the centroid, major and minor axis, 
and average radius of the isophote in question. After careful examination of LRG 
images provided by SDSS, we adopted a conservative range of galaxy flattening, 
$0.35< b/a < 0.85$, for which misidentification, star contamination, and other 
problems are greatly reduced while keeping at the same time the highest number 
of objects and the lowest number of round shaped galaxies.

We have computed photometric redshifts using the ANNz code for all galaxies in 
fields centered on each LRG target within 
$\sim 2.5~\mpc$. We applied k-corrections using the method of Blanton et 
al (2003), version 4.1.

\begin{figure}\includegraphics[width=0.48\textwidth]{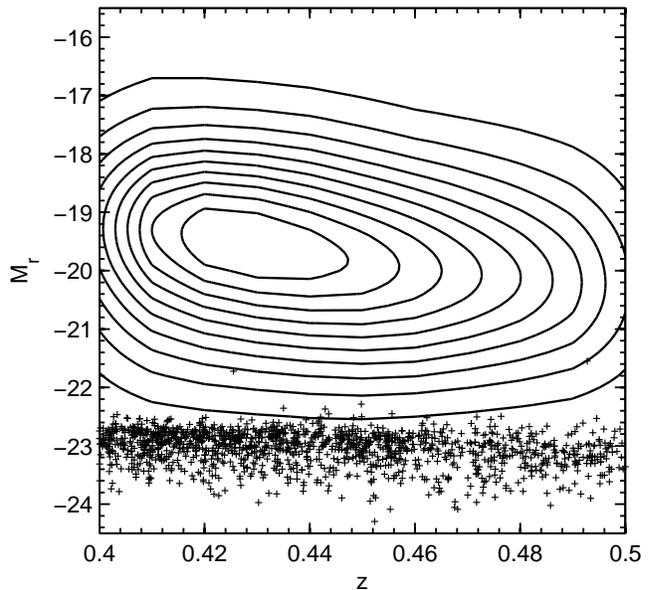}
\caption{Petrosian absolute magnitude $M_r$ vs redshift for $\sim25.000$ tracer
 galaxies with photometrically derived redshifts calculated with the publicity available
 Artificial Neural Network code (Collister \& Lahav (2004), ANNz). The ten contours enclose 
$10\% ~- ~100\%$ respectively. The $+$ signs correspond to the LRG sample.}
\label{figura1}
\end{figure}

In Figure \ref{figura1}, we show the distribution of redshifts and Petrosian 
absolute magnitudes of the spectroscopic LRG target sample, and of $\sim~25.000$ 
tracer galaxies with photometrically derived redshifts. It can be appreciated that 
LRGs are typically brighter by $\sim$3.5 magnitudes than the most luminous 
galaxies in the fields.

\section{Analysis and Results}

To search for alignment signals we calculate the relative angle $\phi$ between the 
position angle of the central luminous red galaxy and the vector pointing to 
each tracer galaxy within $1.5\mpc$ and $dv=1000$ km/s; and then count the number 
of galaxies in angular bins between 0\degr and 90\degr . For the quantification of 
the strength of these alignments we define the relative fraction distribution $f(\phi)$ as
\[
f(\phi)=\frac{N(\phi)-\langle N(\phi) \rangle}{\langle N(\phi) \rangle}
\]
where $N(\phi)$ is the number of galaxies at each angular bin and $\langle N(\phi) 
\rangle$ its  mean value. This allows to effectively measure the fractional excess of 
tracer galaxy counts as a function of $\phi$ with respect to the mean. With this 
definition the absence of alignments would be characterized by a flat distribution 
around the zero mean. On the other hand, an excess of galaxy counts at low values 
of $\phi$ indicates the presence 
of an alignment effect. 

In order to assess the statistical significance of alignments we adopt the 
following procedure:

\begin{table*}
\begin{minipage}{170mm}
\centering
\caption{Definition of samples and alignment statistics}
\label{tabla1}
\begin{tabular}{@{}lcccccccccccc}
\hline
Sample & LRG & \multicolumn{2}{c}{Tracer galaxies}\\
\cline{3-4}
Name & Magnitude & Magnitude & Color & $\sigma$ & $\chi^2$ & $\langle\phi\rangle$ & $P_>(\langle\phi\rangle)$ & $b$ & $P_>(b)$ & $n_{45}$ & $P_>(n_{45})$ & $N_{gal}$\\
\hline
S & $M_r<-21.5$ & $M_r<-14.0$ & no restriction & 0.26 & 2.94 & 44.5 & 0.05 & 0.02 & 0.081 & 1.019 & 0.174 & 9883\\
S0 & $M_r<-21.5$ & $M_r<-14.0$ & $(g-r)>1.0$ & 0.49 & 7.00 & 43.7 & 0.002 & 0.066 & 0.002 & 1.100 & 0.006 & 2834\\
S1 & $M_r<-21.5$ & $M_r<-19.8$ & $(g-r)>1.0$ & 0.65 & 10.2 & 42.9 & 0.000 & 0.109 & 0.001 & 1.179 & 0.000 & 1593\\
S2 & $M_r<-21.5$ & $M_r>-19.8$ & $(g-r)>1.0$ & 0.73 & 0.14 & 44.7 & 0.36 & 0.011 & 0.400 & 1.008 & 0.462 & 1241\\
S3 & $M_r<-21.5$ & $M_r<-19.3$ & $(g-r)<0.2$ & 0.75 & 0.95 & 44.2 & 0.14 & 0.036 & 0.171 & 1.033 & 0.291 & 1210\\
S4 & $M_r<-21.5$ & $M_r>-19.3$ & $(g-r)<0.2$ & 1.01 & 0.05 & 45.2 & 0.42 & -0.022 & 0.350 & 0.923 & 0.845 & 656\\
S5 & $M_r<-22.8$ & $M_r<-14.0$ & $(g-r)>1.0$ & 0.54 & 6.97 & 43.5 & 0.002 & 0.076 & 0.003 & 1.120 & 0.000 & 2241\\
S6 & $M_r>-22.8$ & $M_r<-14.0$ & $(g-r)>1.0$ & 1.06 & 0.42 & 44.3 & 0.26 & 0.031 & 0.293 & 1.031 & 0.365 & 593\\
S7 & $M_r<-22.8$ & $M_r<-14.0$ & $(g-r)<0.2$ & 0.69 & 0.21 & 44.6 & 0.29 & 0.008 & 0.380 & 0.959 & 0.769 & 1391\\
S8 & $M_r>-22.9$ & $M_r<-14.0$ & $(g-r)<0.2$ & 0.95 & 0.08 & 44.7 & 0.39 & 0.011 & 0.410 & 1.043 & 0.279 & 748\\
\hline
\end{tabular}
\end{minipage}
\end{table*}

\begin{enumerate}
\item We compute the mean and dispersion of the relative angle $\phi$ between the 
LRG and the tracers. For an isotropic distribution is expected 
$\langle \phi \rangle  \approx 45$\degr while any departure from this value 
indicates either alignments or anti-alignments.
\item We also calculate the distribution of $\delta = \langle \phi \rangle - 45$\degr
which should exhibit for the isotropic case a mean value 
$\langle \delta \rangle \approx 0$ and standard deviation $\sigma = 90 / 
\sqrt{12N}$, where $N$ is the total number of LBG-galaxy pairs involved (Struble \& Peebles (1985)). 
The mean and dispersion of these deviations across all galaxies in the sample allows us  
to construct  $\chi ^2 = \delta ^2 / \sigma ^2$, which is another useful 
measure of the statistical significance of the results. 
\item We fit the $f(\phi)$ distribution with $f(\phi)=b$ $cos(2\phi)$ and quote $b$ 
values. This coefficient quantifies the anisotropy amplitude 
($b>0$ alignment, $b<0$ anti-alignment).
\item We calculate the ratio between the number of $\phi$ values lesser (grater) 
than 45\degr and define $n_{45}= 2 N_{<45}/ N_{>45}$.
\end{enumerate}
To address the question of how significant are the alignments signals detected we 
proceed as follows. We calculate the distribution of random occurrence of  
$\langle \phi \rangle$, $b$, and $n_{45}$ using a Monte-Carlo method
assigning random position angles to the central LRGs. By computing 
the statistics for 1000 random realizations we provide a robust estimate of 
the reliability of the results independently of a possible non-gaussian behavior. Also, 
since the galaxies used for each random sample are exactly the same as real 
data ones, with the same radial dependence, and the same physical properties; 
we can avoid any bias caused by clustering or other effects, leaving 
the LRG orientation as the only parameter that varies across the random set. 
Then, if an observed parameter value is well outside the corresponding 
distribution of the random realization we assure its reliability since it 
could hardly been obtained by chance. We compute the 
probability of obtaining values higher than the observed ones for $\langle \phi\rangle$, 
$b$ and $n_{45}$, quoted in Table \ref{tabla1} as $P_>(\langle\phi\rangle)$, $P_>(b)$ 
and $P_>(n_{45})$, respectively. Low values of these probabilities indicate a strong 
confidence of the observed quantities being non-random.

\begin{figure}
\includegraphics[width=0.45\textwidth]{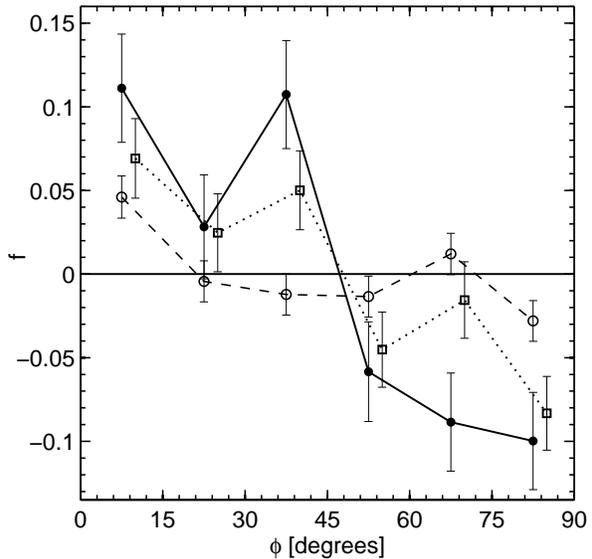}
\caption{
Normalized distribution of the relative angle $\Phi$ between LRGs position 
angle and the radius vector to neighbors within $rp\la 1.5~\mpc$ and 
$dv=1000$ km/s. The dashed lines correspond to $M_r<-14$ 
(sample S), the dotted lines to sample S0 ($M_r<-14.0$ and $(g-r)>1$), 
while the solid lines correspond to $M_r<-19.8$ and $(g-r)>1$ (sample S1). In the 
case of no alignment we expect $f=0$. It can be appreciated a significant alignment effect 
between luminous red galaxies and red neighbors (samples S0 and S1). Error bars 
correspond to Poisson statistical uncertainty.}
\label{figura2}
\end{figure}

\begin{figure*}\includegraphics[width=0.8\textwidth]{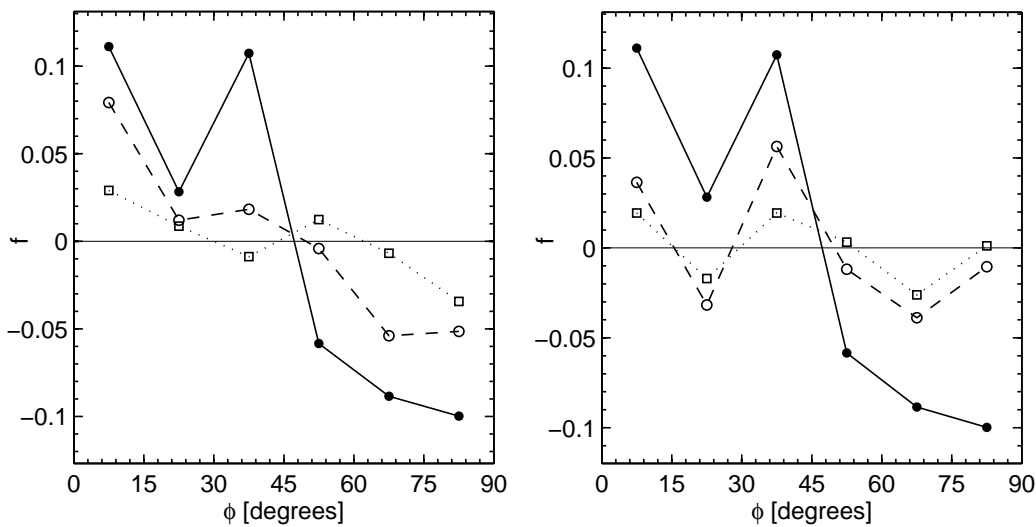}
\caption{
Alignment signal for sample S1 for different values of relative radial velocity 
difference $dv$ between the central LRG and the surrounding tracers (left), 
and for different projected radius $rp$ (right). The panel for $dv$ shows how 
the relative excess of galaxy counts for $dv=1000$ km/s (solid line), $dv=3000$ 
km/s (dashed line) and $dv=10000$  km/s (dotted line) gradually tending to isotropy. 
A similar tendency is shown in the right panel for values of $rp=1.5~\mpc$ (solid line), 
$rp=2.5~\mpc$ (dashed line) and $rp=3~\mpc$ (dotted line).}
\label{figura3}
\end{figure*}

\begin{figure*}\includegraphics[width=0.8\textwidth]{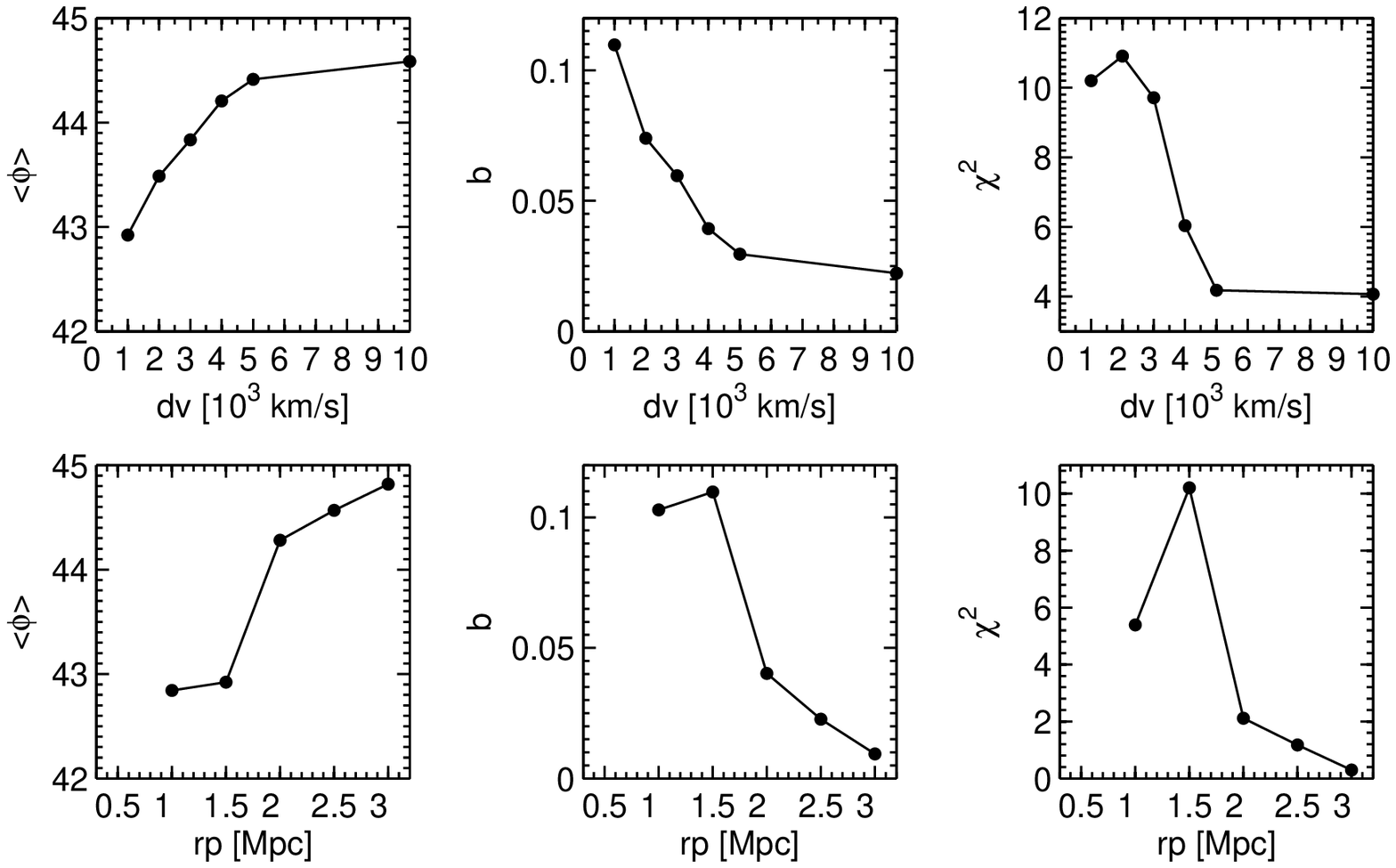}
\caption{
Dependence of $\langle\phi\rangle$, $\chi^2$ and $b$ on velocity difference (top row) 
and projected radius (bottom row) for sample S1. It is clear that for larger separations (in both 
velocity and projected radius) all parameters exhibit a behavior consistent with a gradual 
tendency to isotropy.}
\label{figura4}
\end{figure*}

We applied this set of statistical tests to our sample of LRGs and surrounding galaxies, 
and extracted different subsamples taking into account luminosity and color. 
We perform this analysis taking into account Yang et al. (2006) results, who found the red population to be 
more strongly aligned, an effect that is desired to be tested in the present analysis. 
By considering the tracer galaxy distributions of luminosity, shown in Figure 
\ref{figura1}, and ($g-r$) colour index, we have constructed samples of luminous (faint) 
tracers, $M_r < -19.8$ ($M_r > -19.8$); and red (blue) tracers with $(g-r) > 1$ 
($(g-r) < 0.2$). In addition, we also analyzed complementary samples of bright (faint) 
LRGs, $M_r < -22.8$ ($M_r > -22.8$); with the same red (blue) population of tracers. 
All samples are defined in Table \ref{tabla1}, conveniently labeled, and listed along 
with the obtained parameters of alignment and significance.

The results for samples S, S0 and S1 are shown in figure \ref{figura2}. There is a 
marginal alignment signal for sample S, while sample S0 has a larger and significant 
alignment amplitude. It is also clear that the most significant alignment signal is found for 
bright, red tracers (sample S1). This red bright population is strongly aligned as is evident 
by inspection to table \ref{tabla1}. The magnitude of the effect is of the order 
of $11\%$ as accounted by the amplitude of the cosine fit ($\langle \phi \rangle =42\degr.9$). 
Given the strong alignment detected for bright, red tracers, we have also explored for 
faint red tracers (sample S2) finding no significant alignment signal, 
as can be appreciated in table \ref{tabla1}. So that, the alignment signal for 
sample S0 is likely to be due almost entirely to the bright galaxies of sample S1.

\begin{figure*}\includegraphics[width=0.7\textwidth]{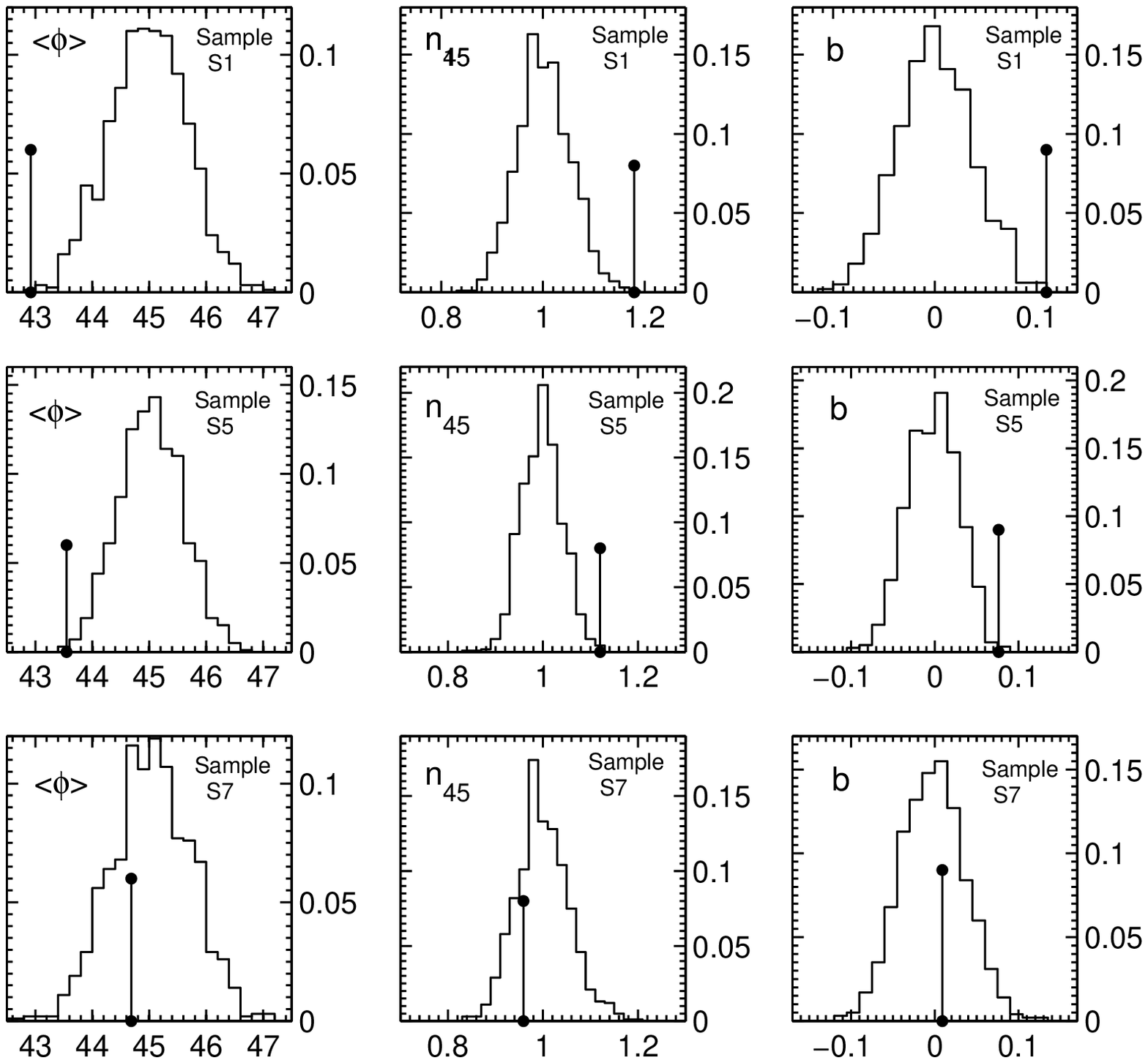}
\caption{
Distribution of statistical parameters $\langle\phi\rangle$, $b$ and $n_{45}$ 
drawn from 1000 samples where the LRG position angle $\theta$ is extracted from 
homogeneous distribution. Vertical lines indicate the values obtained directly from observed data 
for the different samples (see Table \ref{tabla1}). The first row corresponds to the 
bright red population of tracer galaxies around LRGs, showing a significant alignment 
signal singe by observed values are well outside the random distributions. The second 
row (sample S5, bright LRGs, red tracers) also show a correlation. The third row is a 
good example of no alignment effect consistent with isotropy corresponding to blue 
neighbors surrounding faint LRGs (sample S7).
}
\label{figura5}
\end{figure*}

It is natural to expect a dependence between alignment effects and the velocity difference 
$dv$ between central and surrounding tracer galaxies, so that an 
additional analysis is shown in figure \ref{figura3}, where the alignment signal and 
statistical parameters for sample S1 (bright red population) are plotted at several 
intervals of $dv$ in the range $10^3 - 10^4$ km/s. As expected, the alignment 
strength declines with increasing $dv$ values as more uncorrelated 
tracers are included. This tendency is reflected also in the progression of 
$\langle \phi \rangle$ values towards isotropy at 45\degr and the  
lower values of $\chi^2$ and alignment amplitude $b$. For $dv<1000$ km/s, the number 
of objects fall well beyond reasonable limits. Therefore, we adopt this limit value 
as the optimal one for the purposes of this analysis. 

In figure \ref{figura4}, we show the dependence of the alignment signal 
($\langle \phi \rangle , ~ b,$ and $\chi^2$) of sample S1 on relative velocity 
$dv$ and projected separation ($r_p$). As it can be appreciated, there is a 
smooth tendency to isotropy as either $dv$ and $r_p$ increase.

In order to explore if the alignment effects comprising the red population extend to other 
type of galaxies we tested the alignment in two samples with both, bright and faint 
blue tracers (samples S3 and S4, respectively). As it can be appreciated in table 
\ref{tabla1}, the results are fairly consistent with isotropy, implying no correlation 
between LRG orientation and blue late-type neighbors, a fact that is in good 
agreement with most previous investigations about alignments at low redshift 
(Yang et al. (2006)).

The question whether the luminosity of the central LRG affects the alignment 
pattern with  neighboring galaxies is also worth to be investigated. For this 
matter, we repeated the  analysis for samples of bright (faint) LRGs, with 
$M_r < -22.8$ ($M_r > -22.8$), and red tracers with $(g-r) > 1$ (samples S5 
and S6, respectively).  This indicates that the most significant alignment signal is 
obtained for bright LRGs even while considering red tracer galaxies. 
As the results described above suggest that color is a parameter that determines
the presence or lack of correlation with the LRG position angle, we performed the 
test again, but considering blue tracer $(g-r) < 0.2$, with the same magnitude cut 
dividing bright and faint central LRGs (samples S7 and S8). The results indicate 
clearly that these cases are consistent with isotropy. 
Again, blue galaxy positions do not show correlation with LRG 
orientations, independently of their luminosity.

A visualization of the reliability of the results is given in the 
panels of figure \ref{figura5} which show the distributions of occurrence of 
the alignment parameters $\langle \phi \rangle ,~ n_{45}$, and $b$ for the Monte-Carlo 
realizations with randomly assigned LRG's position angles for samples S1, S5 and S7. 
It can be appreciated that true value of the parameters for samples S1 and S5 are well beyond 
the Monte-Carlo simulation results indicating a statistically significant departure 
from isotropy. On the other hand, in sample S7 the observed values are well within the 
Monte-Carlo distributions.

\section{Conclusions}
From the statistical analysis applied to the data we may summarize the 
following results:
\begin{enumerate}
\item A study of alignments equivalent to those typically performed at 
lower redshifts using information spectroscopic Catalogue can 
also be performed at higher redshifts, using distance estimates 
obtained via photometric redshift techniques. This allows to probe 
deeper into the universe, overcoming the natural limitation of wide-area 
redshift surveys. 
\item The orientation of objects from the Luminous Red Galaxy sample 
extracted from the SDSS shows a significant alignment signal with respect 
to the direction to neighbors within $1.5~\mpc$ involving an excess of $11\%$ 
galaxies respect to an uncorrelated population. This signal stands at over 
$3\sigma$ level only for the bright red population of tracer galaxies. 
The probability of obtaining such alignment pattern from 1000 random realizations is 
lesser than $1$.
\item While the above tendency is quite remarkable for red, bright, neighbors,
 we found no evidence of such an effect for bluer and/or fainter galaxies around LRGs.
\end{enumerate}

\label{lastpage}
\end{document}